\documentclass[conference]{IEEEtran}
\IEEEoverridecommandlockouts
\ifCLASSINFOpdf
\else
   \usepackage{graphicx}
   \graphicspath{{../eps/}}
   \DeclareGraphicsExtensions{.eps}
\fi
\usepackage{amsthm,amsmath,amssymb}
\usepackage{mathrsfs}
\usepackage{amsfonts}
\usepackage{graphicx, amssymb}
\usepackage{amsthm}
\usepackage{amsmath}
\usepackage{epsfig}
\usepackage{graphicx}
\usepackage{graphics}
\usepackage{multirow}
\usepackage{threeparttable}
\usepackage{amsthm}
\usepackage{float}
\usepackage{color}
\usepackage{algorithm}
\usepackage{algpseudocode}
\usepackage{bm}
\usepackage{algorithmicx}
\usepackage{algpseudocode}
\usepackage{setspace}
\usepackage{authblk}
\usepackage{cite}
\usepackage{pifont}
\usepackage{xcolor}
\usepackage{epstopdf}
\usepackage{caption}
\usepackage{graphicx}
\usepackage{float}
\usepackage{subcaption}
\usepackage{cleveref}
\theoremstyle{definition}

\newcommand{\RNum}[1]{\uppercase\expandafter{\romannumeral #1\relax}}

\usepackage{wrapfig}
\usepackage{psfrag}
\usepackage{graphicx}
\usepackage{threeparttable}
\usepackage{cases}
\usepackage{subeqnarray}
\usepackage{color}
\usepackage{underscore}

\usepackage{textcomp}
\usepackage{xcolor}
\def\BibTeX{{\rm B\kern-.05em{\sc i\kern-.025em b}\kern-.08em
    T\kern-.1667em\lower.7ex\hbox{E}\kern-.125emX}}

\begin{document}
\title{Little Pilot is Needed for Channel Estimation with Integrated Super-Resolution Sensing and Communication}

\author{{Jingran~Xu{*}, Huizhi Wang{*}, Yong~Zeng{*$\dagger $}, and Xiaoli Xu{*}}\\
{{*}National~Mobile~Communications~Research~Laboratory,~Southeast~University,~Nanjing~210096,~China}\\
{{$\dagger $}Purple~Mountain~Laboratories,~Nanjing~211111,~China}\\
Email:~{\{jingran_xu,~wanghuizhi,~yong_zeng, xiaolixu\}}@seu.edu.cn
}

%
%
\maketitle

\begin{abstract}
Integrated super-resolution sensing and communication (ISSAC)
is a promising technology to achieve extremely high sensing performance for critical parameters, such as the angles of the wireless channels.
In this paper, we propose an ISSAC-based channel estimation method, which requires little or even no pilot, yet still achieves accurate channel state information (CSI) estimation. The key idea is to exploit the fact that subspace-based super-resolution algorithms such as multiple signal classification (MUSIC) do not require a priori known pilots for accurate parameter estimation. Therefore, in the proposed method, the angles of the multi-path channel components are first estimated in a pilot-free manner while communication data symbols are sent. After that, the multi-path channel coefficients are estimated, where very little pilots are needed. The reasons are two folds. First, compared to the conventional channel estimation methods purely relying on channel training, much fewer parameters need to be estimated once the multi-path angles are accurately estimated. Besides, with angles obtained, the beamforming gain is also enjoyed when pilots are sent to estimate the channel path gains. To rigorously study the performance of the proposed method, we first consider the basic line-of-sight (LoS) channel.
By analyzing the minimum mean square error (MMSE) of channel estimation
and the resulting beamforming gains,
we show that our proposed method significantly outperforms the conventional methods purely based on channel training.
We then extend the study to the more general multipath channels.
Simulation results are provided to demonstrate our theoretical results.
\end{abstract}

\IEEEpeerreviewmaketitle
\section{Introduction}
%
%
%

Integrated sensing and communication (ISAC) has been identified as one of the six usage scenarios for
IMT-2030 (6G) \cite{IMT2030}.
In particular, the concept of integrated super-resolution sensing and communication (ISSAC) is further proposed \cite{zhangchaoyue}, which uses super-resolution algorithms in ISAC systems
to achieve extremely high sensing performance for those critical parameters, such as delay, Doppler, and angle of the sensing targets.
Furthermore, due to the spatial sparsity,
high-frequency channels such as millimeter wave (mmWave) and TeraHertz (THz)
are dominated by only a small number
of multipath components. Therefore, those channels can be
modeled in a parametric form, in terms of the path angles
of departure/arrival (AoD/AoA) and the corresponding path
gains \cite{mmwavechannel1}. As a result,
the mm-wave or THz channel estimation problem can be transformed into the estimation of path directions and channel gains, rather than the estimation of the entire multiple input multiple output (MIMO) channel matrix.


For conventional pilot-based channel estimation methods \cite{estimationmethod}, in order to obtain an accurate channel state information (CSI), least squares (LS) and linear minimum mean square error (LMMSE) methods are usually used by scrutinizing the relationship between the received signals and known pilot sequences.
Maximum likelihood estimation (MLE) employs a likelihood-based approach to maximize the probability of observed signals given the pilot sequence.
However, massive MIMO will result in increased pilot symbols
and high computational complexity for conventional pilot estimation method due to the substantial number of antennas, thereby impacting spectral efficiency and real-time processing demands.
Channel sparsity together with various compressive sensing (CS) algorithms
are used to reduce the training overhead for massive MIMO systems \cite{CS1}.
However, the underlying requirement of the high signal-to-noise ratio (SNR),
an appropriate dictionary matrix, and perfect measurement feedback for users to
base station (BS) is a matter of concern.
An alternative low-rank model for massive uniform linear array (ULA)
are proposed \cite{gaofeifei1}, which exploits angular information via array signal processing.
Techniques such as discrete Fourier
transformation (DFT) \cite{DFT}, multiple signal
classification (MUSIC) algorithm \cite{music1} and estimation of
signal parameters via rotational invariance technique (ESPRIT) using the subspaces estimated by matrix or tensor decompositions \cite{ESPRIT},
have been leveraged to estimate the unknown
parameters in angular channel models.

The conventional channel estimation methods that using angle information rely on transmitting pilot signals \cite{DFT,music1,ESPRIT}, and thus abundant pilot resources are still needed.
In this paper, we propose an ISSAC-based channel estimation method, which requires little or even no
pilot, yet still achieves accurate CSI estimation.
The key idea is that subspace-based
super-resolution algorithms, such as MUSIC, can accurately estimate
parameters without using a priori known pilots.
Therefore, we can first accurately estimate the angles of the multi-path channel components
during the phase of data transmission without using any pilot.
After that, by exploiting the beamforming gain with angles obtained,
the multi-path channel coefficients are estimated,
where very little pilots are needed with much fewer parameters
compared to the conventional channel estimation methods.
To rigorously study the performance of the proposed
method, we first analyze the MMSE of channel
estimation and the resulting beamforming gains in the basic line of sight (LoS) channel.
Then the study is extended to the more general multipath channels.
Simulation results are provided to demonstrate that
our proposed method significantly outperforms the conventional
methods purely based on channel training.
\section{System Model}
\begin{figure}[!t]
 \setlength{\belowcaptionskip}{-0.2cm}
  \centering
  \centerline{\includegraphics[width=2.5in,height=1.3in]{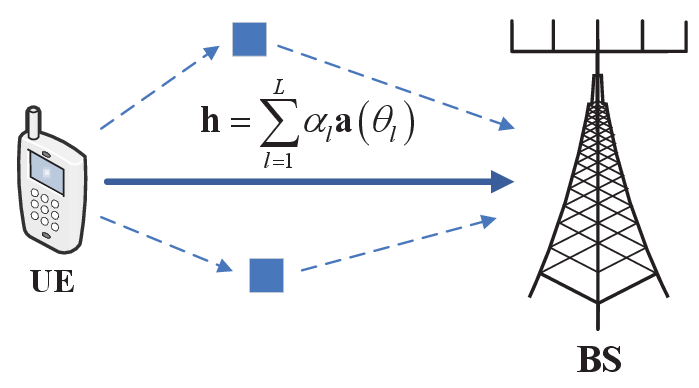}}
  \caption{A SIMO system, where the channel from the UE to the BS includes $L$ paths.}
  \label{SIMOsystem model}
  \vspace{-0.3cm}
  \end{figure}
As shown in Fig. \ref{SIMOsystem model}, we consider an uplink mmWave communication system, with a single-antenna user equipment (UE) communicating with a BS equipped with $M\gg 1$ antennas.
For the conventional channel estimation scheme \cite{pilotestimation}, the symbol transmission is divided into two phases during each channel coherence time, where the pilot transmission period is needed first to estimate the channel. In particular, denote by $\rho $ and $\kappa $ the length of uplink training and data transmission sequence, respectively.
Let ${P}_{t}$ and ${P}_{d}$ denote the transmit power by the UE in pilot and data transmission phases, respectively.
Further denote by ${\varphi}(n)$ the pilot sequence sent by the UE with $\sum\limits_{i=1}^{\rho }{{{\left| \varphi \left( i \right) \right|}^{2}}}=\rho $, which is known to the receiver.
Then the received signals by the BS during the pilot symbol durations can be written as
\begin{equation}
	{{{\mathbf{y}}_{t}(n)}=\sqrt{{{P}_{t}}}\mathbf{h}{{\varphi }\left( n \right)}+{{\mathbf{z}}_{t}(n)},n=1,...,\rho}. \label{originalpilotphase}
\end{equation}
Denote the i.i.d. information-bearing symbols of UE as $s(n)$,
which follows circularly symmetric complex
Gaussian (CSCG) distribution with normalized power, i.e.,
$s(n)\sim\mathcal{C}\mathcal{N}\left( 0, 1\right)$.
Then the received signals during data symbol durations can be written as
\begin{equation}
	{{{\mathbf{y}}_{d}(n)}=\sqrt{{{P}_{d}}}\mathbf{h}{s}\left( n \right)+{\mathbf{z}}_{d}(n)},n=\rho+1,...,\rho+\kappa, \label{originaldataphase}
\end{equation}
where ${{\mathbf{z}}_{t}(n)}$ and ${{\mathbf{z}}_{d}(n)}$ denote the i.i.d. CSCG noise with zero-mean and
variance ${{\sigma }^{2}}$.
Further we denote the receive signal-to-noise ratio (SNR) of the pilot and data transmission at the BS without beamforming as ${\mathrm{SNR}}_{t}$ and ${\mathrm{SNR}}_{d}$, respectively,
i.e., ${\mathrm{SNR}}_{t}=\frac{{{P}_{t}}}{M\sigma^{2}}{{\left\| \mathbf{h} \right\|}^{2}}$,
${\mathrm{SNR}}_{d}=\frac{{{P}_{d}}}{M\sigma^{2}}{{\left\| \mathbf{h} \right\|}^{2}}$.

With the pilot symbol ${\varphi }\left( n \right)$ known at the BS, the LS estimation of $\mathbf h$ can be written as \cite{pilotestimation}
\begin{equation}
	{{\bf{\hat h}}}_{\rm{cp}}=\frac{1}{\sqrt{{{P}_{t}}{{\rho }^{2}}}}\sum\limits_{i=1}^{\rho }{{{\mathbf{y}}_{t}}\left( i \right){{\varphi }^{*}}\left( i \right)}=\mathbf{h}+\frac{1}{\sqrt{{{P}_{t}}{{\tau }^{2}}}}\mathbf{z}_{t}^{\prime },
\end{equation}
where $\mathbf{z}_{t}^{\prime }=\sum\limits_{i=1}^{\rho }{{{\mathbf{z}}_{t}}\left( i \right)}{{\varphi }^{*}}\left( i \right)$ is the resulting noise vector. It can be
shown that ${\mathbf{z}}_{t}^{\prime }$ is i.i.d. CSCG noise with power ${{\rho }}{{\sigma }^{2}}$, i.e.,
${\mathbf{z}}_{t}^{\prime }\sim\mathcal{C}\mathcal{N}\left( \mathbf 0, {{\rho }}{{\sigma }^{2}}{\mathbf{I}}_{M}\right)$.

To detect the communication signal, a receive beamforming vector $\mathbf{v}_{\mathrm{cp}}=\frac{{\mathbf{\hat{h}_{\mathrm{cp}}}}}{\left\| {\mathbf{\hat{h}_{\mathrm{cp}}}} \right\|}$ is used.
Then the resulting signal is written as
\begin{equation}
	{{{y}_{d,\mathrm{cp}}}(n)=\mathbf{v}_{\mathrm{cp}}^{\mathrm{H}}{{\mathbf{y}}_{d}(n)}
		=\sqrt{{{P}_{d}}}\mathbf{v}_{\mathrm{cp}}^{\mathrm{H}}\mathbf{h}{s}\left( n \right)+z_{d,\mathrm{cp}}\left( n \right)},\label{afterbeampilot}
\end{equation}
where $z_{d,\mathrm{cp}}\left( n \right)=\mathbf{v}_{\mathrm{cp}}^{\mathrm{H}}{{\mathbf{z}}_{d}(n)}$ denotes the CSCG noise with zero-mean and variance ${{\sigma }^{2}}$.
The expected SNR of the signal in \eqref{afterbeampilot} is
\begin{equation}
	{{{\gamma }_{\mathrm{cp}}}=\mathbb{E}\Big[ \frac{{{P}_{d}}{{\left| \mathbf{v}_{\mathrm{cp}}^{\mathrm{H}}\mathbf{h} \right|}^{2}}}{{{\sigma }^{2}}} \Big]
		=\frac{{{P}_{d}}}{{{\sigma }^{2}}}{{\mathbf{h}}^{\mathrm{H}}}\mathbb{E}\left[ {{\mathbf{v}}_{\mathrm{cp}}}\mathbf{v}_{\mathrm{cp}}^{\mathrm{H}} \right]\mathbf{h}}.\label{gamma1initial}
\end{equation}

\emph{Theorem 1:}
The expected SNR of the signal in \eqref{gamma1initial} can be further approximated as
\begin{equation}
\setlength\abovedisplayskip{1pt}
\setlength\belowdisplayskip{1pt}
	{{{\gamma }_{\mathrm{cp}}}\approx \frac{{{P}_{d}}{{\left\| \mathbf{h} \right\|}^{2}}}{{{\sigma }^{2}}}\left( 1-\xi \right)},\label{gamma1app}
\end{equation}
where $\xi=\frac{1-\frac{1}{M}}{\rho {\mathrm{SNR}}_{t}+1}$ is the penalty loss caused by imperfectly estimated channel.
\begin{IEEEproof}
	Please refer to Appendix A.
\end{IEEEproof}
It is noted that $\xi$ increases with $M$, which undermines the performance of data transmission.
This is because the beamforming gain is not exploited during the channel training phase.
Furthermore, it is shown in \eqref{gamma1app} that the expected SNR ${{\gamma }_{\mathrm{cp}}}$ increases monotonically
with $\rho \mathrm{SNR}_{t}$, which can be defined as the pilot energy consumption.
There exists an upper bound for ${\gamma }_{\mathrm{cp}}$ as
\begin{equation}
\setlength\abovedisplayskip{1pt}
\setlength\belowdisplayskip{1pt}
	{{{\gamma }_{\mathrm{cp}}}<{{\gamma }_{\mathrm{cp},\mathrm{upper}}}=\frac{{{P}_{d}}{{\left\| \mathbf{h} \right\|}^{2}}}{{{\sigma }^{2}}}},
	\label{upper}
\end{equation}
which can be reached when $\rho \mathrm{SNR}_{t}\to \infty$.


On the other hand, it is obvious that the estimation accuracy can be improved for longer pilot length by calculating the minimum mean square error (MMSE) of $\mathbf{h}$ as
\begin{equation}
\setlength\abovedisplayskip{1pt}
\setlength\belowdisplayskip{1pt}
	{e_{{\rm{cp}}}} = \mathbb{E}\Big[ {{{\Big\| {{{{\bf{\hat h}}}_{{\rm{cp}}}} - {\bf{h}}} \Big\|}^2}} \Big] = \frac{{M{\sigma ^2}}}{{{P_t}\rho }}.\label{RMSEe1}
\end{equation}

Therefore, it seems that we need to make the pilot sufficient long, in order to obtain accurate channel estimation results and higher SNR. However, this may introduce overhead and reduce the
spectral efficiency. In this paper, we ask the following question: Is it possible to use little amount of pilots while still guarantee the communication performance?
In order to answer this question, we first introduce the channel model in the parametric form, in terms of the path angles
of AoD/AoA and the corresponding path gains in \cite{mmwavechannel1}.

Specifically, we let $\mathbf{h}=\sum\limits_{l=1}^{L}{{{\alpha }_{l}}\mathbf{a}\left( {{\theta }_{l}} \right)}$ denote the channel from the UE to the BS, where $L$ is the number of multipaths,
${{\alpha }_{l}}$ denotes the complex-valued path
coefficient, and $\mathbf{a}\left( {{\theta }_{l}} \right)$ denotes the BS array response vector. For ULA with half-wavelength spacing, $\mathbf{a}\left( {{\theta }_{l}} \right)={{\left[ 1,{{e}^{j\pi\sin \left( {{\theta }_{l}} \right)}},\cdots ,{{e}^{j\pi\left( M-1 \right)\sin \left( {{\theta }_{l}} \right)}} \right]}^{\mathrm{T}}}\in {{\mathbb{C}}^{M\times 1}}$.
Let $\Theta=\left({{\theta }_{1}},\cdots ,{{\theta }_{L}} \right)$, $\mathbf{A}\left(\Theta  \right)=\left[ \mathbf{a}\left( {{\theta }_{1}} \right),\cdots ,\mathbf{a}\left( {{\theta }_{L}} \right) \right]$, $\bm{ \alpha}={{\left[ {{\alpha }_{1}},\cdots ,{{\alpha }_{L}} \right]}^{\mathrm{T}}}$,
and $\mathbf{h}$ can also be written as
\begin{equation}
\setlength\abovedisplayskip{1pt}
\setlength\belowdisplayskip{1pt}
{\mathbf{h}=\mathbf{A\left(\Theta  \right)}\bm{ \alpha}}\label{channelh}.
\end{equation}
Typically, the channel estimation process include the estimation of the angle of arrivals $\bf {\Theta}$ (AoAs) and the complex-valued path coefficient $\bm {\alpha}$. It is noted that $\bf {\Theta}$ and $\bm {\alpha}$ alter with different timescales.
Specifically, $\bf{\Theta}$ usually changes much slower than the complex-valued path coefficient $\bm {\alpha}$, which can be assumed unchanged during each path-invariant block \cite{pathinvariant1,pathinvariant3,pathinvariant4}.
On the other hand, $\bm {\alpha}$  varies between different channel coherence time.
\section{ISSAC-based Channel Estimation}
In this section, we propose an ISSAC-based channel estimation method where little or even no pilot is needed.
\subsection{Characteristics of the Proposed Method}
Different from prior works in angle-aided channel estimation \cite{DFT,music1,ESPRIT}, which are based on dedicated pilot signals.
In our proposed method, both the existing pilot and data symbols can be utilized to estimate the angles of the multi-path channel components based on subspace-based super-resolution algorithms.
Furthermore, based on the fact that the path AoA $\bm \Theta$ remains unchanged during several coherence time durations, the estimation result in the former coherence blocks is still valid for the following ones.
In addition, with angles obtained during each path-invariant block, the beamforming gain
can be enjoyed when pilots are sent to estimate the channel path gains $\bm \alpha$, which can further improve the accuracy of channel estimation.

\begin{figure}[!t]
 \setlength{\belowcaptionskip}{-0.1cm}
  \centering
  \centerline{\includegraphics[width=3.5in,height=1.9in]{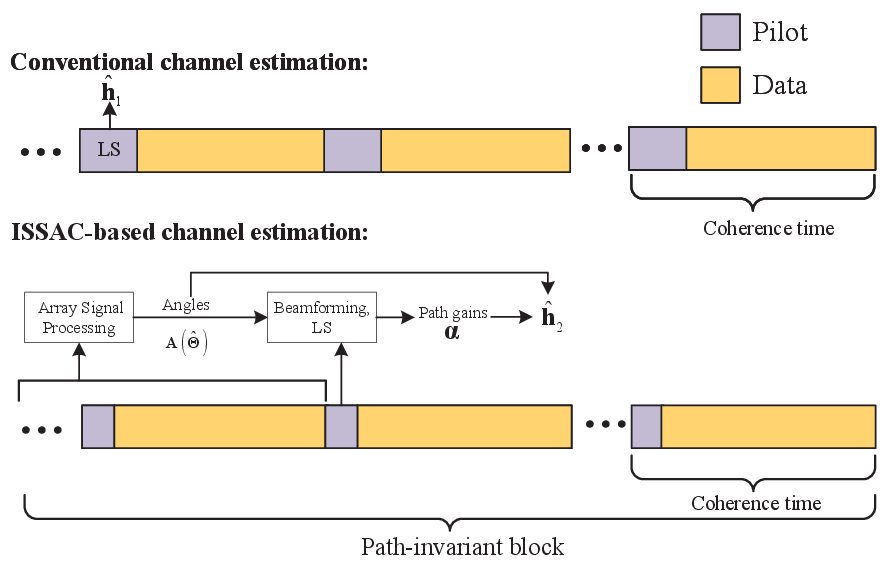}}
  \caption{Comparison between the conventional channel estimation method and the ISSAC-based channel estimation method with little pilot.}
  \label{comparison}
  \vspace{-0.3cm}
\end{figure}

Specifically, the proposed channel estimation scheme involves two stages, as shown in Fig. \ref{comparison}. First, the BS estimates the path directions based on classical sensing algorithms such as the periodogram \cite{periodogram1} algorithm, spectral-based algorithms such as MUSIC \cite{musicDEDI} or ESPRIT \cite{ESPRITDEDI}.
Then, the BS utilizes little pilot and performs the receive beamforming by matching the path directions obtained in previous stage to estimate the complex-valued path coefficients.
\subsection{Performance Analysis}
Next we elaborate the performance of the proposed channel estimation based on ISSAC for channel estimation and signal decoding.
\subsubsection{Free-space LoS Channel}\label{los}
In this case, the channel is dominated by the LoS path and hence we can assume $L=1$, and channel vector can be written as ${\bf h}=\alpha \mathbf{a}\left( \theta  \right)$.
To exploit the data from multiple block shots, we use the Bartlett estimation method, which is a power spectra estimation method that is able to reduce the variance of the periodogram \cite{angleestimation}.

Denote the covariance of $\mathbf{y}_{t}(n)$ in \eqref{originalpilotphase} and $\mathbf{y}_{d}(n)$ in \eqref{originaldataphase} as $\mathbf{R}_{y}$,
which is approximated by its sample average of the received signal with its conjugate transpose
\begin{equation}
{{{\mathbf{R}}_{y}}\approx\frac{1}{\rho +\kappa }(\sum\limits_{i=1}^{\rho }{{{\mathbf{y}}_{t}}(i)\mathbf{y}_{t}^{\mathrm{H}}(i)}+\sum\limits_{i=1}^{\kappa }{{{\mathbf{y}}_{d}}(i)\mathbf{y}_{d}^{\mathrm{H}}(i)})}.
\end{equation}
By performing the 1-D spectral search of the Bartlett pseudospectrum ${{{P}_{\mathrm{Bartlett}}}\left( \theta  \right)={{\mathbf{a}}^{\mathrm{H}}}\left( \theta  \right){{\mathbf{R}}_{y}}\mathbf{a}\left( \theta  \right)}$, the estimation of angle $\hat{\theta }$ can be obtained.

After estimating the angle $\hat{\theta }$, the receive beamforming vector ${{\mathbf{v}}_{\mathrm{lp}}}=\frac{\mathbf{a}\left( {\hat{\theta }} \right)}{\left\| \mathbf{a}\left( {\hat{\theta }} \right) \right\|}$ can be used at the BS to detect communication symbols, and the resulting signal is
	\begin{equation}
		{{y_{d,\mathrm{lp}}(n)}=\mathbf{v}_{\mathrm{lp}}^{\mathrm{H}}{{\mathbf{y}}_{d}(n)}=\sqrt{{{P}_{d}}}\mathbf{v}_{\mathrm{lp}}^{\mathrm{H}}\mathbf{h}{s}\left( n \right)+{z}_{d,\mathrm{lp}}\left( n \right)},\label{afterbeamangleL1}
	\end{equation}
where ${z}_{d,\mathrm{lp}}\left( n \right)=\mathbf{v}_{\mathrm{lp}}^{\mathrm{H}}{{\mathbf{z}}_{d}\left( n \right)}$ denotes the CSCG noise with zero-mean and variance ${\sigma }^{2}$.

\emph{Theorem 2:}
The expected SNR of the signal in \eqref{afterbeamangleL1} can be written as
	\begin{equation}
		\small
		\begin{split}
			\setlength\abovedisplayskip{2pt}
			\setlength\belowdisplayskip{2pt}
			 {{\gamma }_{\mathrm{lp}}}=\mathbb{E}\Big[ \frac{{{P}_{d}}{{\left| \mathbf{v}_{\mathrm{lp}}^{\mathrm{H}}{\bf h} \right|}^{2}}}{{{\sigma }^{2}}} \Big]=\frac{{{P}_{d}}{{\left\| {\bf h} \right\|}^{2}}}{{{\sigma }^{2}}}\frac{{{\left| {{\mathbf{a}}^{\mathrm{H}}}\big( {\hat{\theta }} \big)\mathbf{a}\left( \theta \right) \right|}^{2}}}{{{\big\| \mathbf{a}\big( {\hat{\theta }} \big) \big\|}^{2}}{{\left\| \mathbf{a}\left( \theta \right) \right\|}^{2}}}	.\label{gamma2initial}
		\end{split}
	\end{equation}
\begin{IEEEproof}
By substituting ${{\mathbf{v}}_{\mathrm{lp}}}=\frac{\mathbf{a}\left( {\hat{\theta }} \right)}{\sqrt{M}}$ into (\ref{gamma2initial}), we have ${{\gamma }_{\mathrm{lp}}}=\frac{{{P}_{d}}}{M{{\sigma }^{2}}}{{\left| {{\mathbf{a}}^{\mathrm{H}}}\big( {\hat{\theta }} \big){\bf h}_{\mathrm{lp}} \right|}^{2}}$, and Theorem 2 can be obtained.
\end{IEEEproof}
Theorem 2 shows that the expected receive SNR depends on the accuracy of the angle estimation. When the angle is perfectly estimated, i.e., ${\hat{\theta }} \approx \theta$, we have ${{\gamma }_{\mathrm{lp}}}\approx\frac{{{P}_{d}}{{\left\| \mathbf{h} \right\|}^{2}}}{{{\sigma }^{2}}}.$ Therefore, when the angle estimation is accurate, the expected SNR can reach the upper bound ${{\gamma }_{\mathrm{cp},\mathrm{upper}}}$ in (\ref{upper}).

In order to further demodulate the communication symbol, the complex-valued path coefficients needs to be estimated. In this case, little pilot signal is used to perform the receive beamforming towards $\hat{\theta }$ as
\begin{equation}
	\small
\setlength\abovedisplayskip{1pt}
\setlength\belowdisplayskip{1pt}
{y_{t}(n)=\frac{{{\mathbf{a}}^{\mathrm{H}}}\big( {\hat{\theta }} \big)}{\big\| \mathbf{a}\big( {\hat{\theta }} \big) \big\|}{{\mathbf{y}_{t}(n)}}=\sqrt{{{P}_{t}}}\alpha \frac{{{\mathbf{a}}^{\mathrm{H}}}\big( {\hat{\theta }} \big)}{\big\| \mathbf{a}\big( {\hat{\theta }} \big) \big\|}\mathbf{a}\left( \theta  \right)\varphi(n)+z_{t}(n)},
\end{equation}
where $z_{t}(n)=\frac{{{\mathbf{a}}^{\mathrm{H}}}\left( {\hat{\theta }} \right)}{\left\| \mathbf{a}\left( {\hat{\theta }} \right) \right\|}\mathbf{z}_{t}(n).$ Since the pilot symbol $\varphi(n)$ is known, $y_{t}(n)$ can be projected to $\varphi^{*}(n)$, which gives
\begin{equation}
	\small
\setlength\abovedisplayskip{1pt}
\setlength\belowdisplayskip{1pt}
{{{y}_{t}}^{\prime }=\frac{1}{\sqrt{{{P}_{t}}{{\rho }^{2}}}}\sum\limits_{i=1}^{\rho }{{{y}_{t}}\left( i \right)\varphi^{*} \left( i \right)}=\alpha \frac{{{\mathbf{a}}^{\mathrm{H}}}\big( {\hat{\theta }} \big)}{\big\| \mathbf{a}\big( {\hat{\theta }} \big) \big\|}\mathbf{a}\left( \theta  \right)+\frac{1}{\sqrt{{{P}_{t}}{{\rho }^{2}}}}{{z}_{t}}^{\prime }},
\end{equation}
where $z_{t}^{\prime }=\sum\limits_{i=1}^{\rho }{{{z}_{t}}\left( n \right)}{{\varphi }^{*}}\left( n \right)$ is the resulting noise vector, with
$z_{t}^{\prime }\sim\mathcal{C}\mathcal{N}\left( 0, {{\rho }}{{\sigma }^{2}}\right)$.
The LS estimation of $\alpha$ can be written as
\begin{equation}
	\small
\setlength\abovedisplayskip{1pt}
\setlength\belowdisplayskip{1pt}
\hat \alpha  = \frac{{\left\| {{\bf{a}}(\hat \theta )} \right\|{y_t}^\prime }}{{{{\bf{a}}^{\rm{H}}}(\hat \theta ){\bf{a}}\left( \theta  \right)}}\mathop  \approx \limits^{(a)} \frac{{{y_t}^\prime }}{{\left\| {{\bf{a}}(\hat \theta )} \right\|}} = \frac{{{y_t}^\prime }}{{\sqrt M }},\label{alphaes}
\end{equation}
where $(a)$ holds by assuming ${\hat{\theta }} \approx \theta$.
Furthermore, the estimation error of ${\mathbf{h}}$ can be written as
\begin{equation}
	\small
\begin{split}
\setlength\abovedisplayskip{1pt}
\setlength\belowdisplayskip{1pt}
  {{{\mathbf{\tilde{h}}}}_{\mathrm{lp}}}&=\hat{\alpha }\mathbf{a}\big( {\hat{\theta }} \big)-\alpha \mathbf{a}\left( \theta  \right)  =\frac{{{y}_{t}}^{\prime }}{\sqrt{M}}\mathbf{a}\big( {\hat{\theta }} \big)-\alpha \mathbf{a}\left( \theta  \right) \\
 & =\frac{\alpha }{M}{{\mathbf{a}}^{\mathrm{H}}}\big( {\hat{\theta }} \big)\mathbf{a}\left( \theta  \right)\mathbf{a}\big( {\hat{\theta }} \big)-\alpha \mathbf{a}\left( \theta  \right)+\frac{1}{\sqrt{{{P}_{t}}{{\rho }^{2}}M}}{{z}_{t}}^{\prime }\mathbf{a}\big( {\hat{\theta }} \big)\\
 & \mathop  \approx \limits^{(a)}  \frac{1}{\sqrt{{{P}_{t}}{{\rho }^{2}}M}}{{z}_{t}}^{\prime }\mathbf{a}\big( {\hat{\theta }} \big).\label{tildeh2}
\end{split}
\end{equation}
Therefore, the MMSE of channel ${\mathbf{h}}$ can be accordingly written as
\begin{equation}
		\small
\begin{split}
\setlength\abovedisplayskip{1pt}
\setlength\belowdisplayskip{1pt}
& e_\mathrm{lp}=\mathbb{E}\big[ {{\big\| {{{\mathbf{\tilde{h}}}}_{\mathrm{lp}}} \big\|}^{2}} \big]={{\alpha }^{2}}M\!-\!\frac{{{\alpha }^{2}}}{M}{{\left| {{\mathbf{a}}^{\mathrm{H}}}\left( \theta  \right)\mathbf{a}\big( {\hat{\theta }} \big) \right|}^{2}}\!+\!\frac{{{\sigma }^{2}}}{{{P}_{t}}\rho } \mathop  \approx \limits^{(a)}\frac{{{\sigma }^{2}}}{{{P}_{t}}\rho }.\label{RMSEe2}
\end{split}
\end{equation}
Compared with $e_\mathrm{cp}$ in \eqref{RMSEe1}, it can be shown that the MMSE of the channel estimation we proposed is $M$ times smaller than that using pilot-consuming channel estimation method.

\subsubsection{Multipath Channel}
Next we further consider the more general case with $L>1$.
Since the MUSIC algorithm can operate efficiently only if the $L$ propagation paths are independent,
we use the spatial smoothing technique to handle this problem \cite{FBSS}.

Assume that the antenna array of the BS is divided into $P$ subarrays that overlap each other and the number of elements of each subarray is $M_{\mathrm{sub}}=M-P+1$.
Then the output of the $p$-th forward subarray during pilot transmission in \eqref{originalpilotphase} can be written as
\begin{equation}
\begin{split}
\setlength\abovedisplayskip{1pt}
\setlength\belowdisplayskip{1pt}
  {{\mathbf{y}}_{t,p}(n)}& =\left[ {y_{t,p}(n)},\cdots ,{y_{t,\left( p+{{M}_{\mathrm{sub}}}-1 \right)}}(n) \right]^{\mathrm{T}}\\
  & ={{\mathbf{A}}_{p}\left(\Theta  \right)}\bm{\alpha}\sqrt{{{P}_{t}}}{{\varphi}(n)}+\mathbf{z}_{t,p}(n),\label{subequation}
\end{split}
\end{equation}
where ${y_{t,p}(n)}$ is the $p$-th element of ${\mathbf{y}_{t}(n)}$.
${{\mathbf{z}}_{t,p}(n)}=\left[ {z}_{t,p}(n),\cdots ,{z_{t,\left( p+{{M}_{\mathrm{sub}}}-1 \right)}(n)} \right]^{\mathrm{T}}\in {{\mathbb{C}}^{{{M}_{\mathrm{sub}}}\times 1}}$, $p=1,...,P$.
${{y}_{t,m}(n)}$ and ${{z}_{t,m}(n)}, m=1,...,M$ is the $m$ element of ${\mathbf{y}}_{t}(n)$ and ${\mathbf{z}}_{t}(n)$, respectively.
Furthermore, we have ${{\mathbf{A}}_{p}\left(\Theta  \right)}=\left[ {{\mathbf{a}}_{p}}\left( {{\theta }_{1}} \right),\cdots ,{{\mathbf{a}}_{p}}\left( {{\theta }_{L}} \right) \right]\in {{\mathbb{C}}^{{{M}_{\mathrm{sub}}}\times L}}$,
where ${{\mathbf{a}}_{p}}\left( \theta_{l}  \right)={{\left[ {{e}^{j\frac{2\pi }{\lambda }dp\sin \left( \theta_{l}  \right)}},\cdots ,{{e}^{j\frac{2\pi }{\lambda }d\left( p+{{M}_{\mathrm{sub}}}-1 \right)\sin \left( \theta_{l}  \right)}} \right]}^{\mathrm{T}}}\in {{\mathbb{C}}^{{{M}_{\mathrm{sub}}}\times 1}}$.
Let ${\mathbf{D}=\mathrm{diag}\left[ {{e}^{-j\frac{2\pi }{\lambda }d\sin {{\theta }_{1}}}},\cdots ,{{e}^{-j\frac{2\pi }{\lambda }d\sin {{\theta }_{L}}}} \right]},$ and \eqref{subequation} can be expressed as
\begin{equation}
{{{\mathbf{y}}_{t,p}(n)}={{\mathbf{A}}_{1}\left(\Theta  \right)}{{\mathbf{D}}^{p-1}}\bm{\alpha}\sqrt{{{P}_{t}}}{{\varphi}(n)}+\mathbf{z}_{t,p}(n).}\label{y1p}
\end{equation}
Similarly, the output of the $p$-th forward subarray during data transmission in \eqref{originaldataphase} can be written as
\begin{equation}
{{{\mathbf{y}}_{d,p}(n)}={{\mathbf{A}}_{1}\left(\Theta  \right)}{{\mathbf{D}}^{p-1}}\bm{\alpha}\sqrt{{{P}_{d}}}{s(n)}+\mathbf{z}_{d,p}(n).}\label{y2p}
\end{equation}
Denote the covariance of $p$-th subarray in \eqref{y1p} and \eqref{y2p} as $\mathbf{R}_{p}$,
which is approximated by its sample average of the received signal with its conjugate transpose
\begin{equation}
\setlength\abovedisplayskip{1pt}
\setlength\belowdisplayskip{1pt}
{{{\mathbf{R}}_{p}}\approx\frac{1}{\rho +\kappa }(\sum\limits_{i=1}^{\rho }{{{\mathbf{y}}_{t,p}}\left( i \right)\mathbf{y}_{t,p}^{\mathrm{H}}\left( i \right)}+\sum\limits_{i=1}^{\kappa }{{{\mathbf{y}}_{d,p}}\left( i \right)\mathbf{y}_{d,p}^{\mathrm{H}}\left( i \right)})}.
\end{equation}
By averaging the covariance matrices of the $P$ subarrays and using $M_{\mathrm{sub}}$-order exchange matrix,
the bidirectional spatial smoothing covariance matrix can be obtained \cite{FBSS}.
When the number of
elements of each subarray ${{M}_{\mathrm{sub}}}$ satisfies ${{M}_{\mathrm{sub}}} \ge L+1$,
and the number of subarrays $P$ satisfies $2P\ge L$,
the bidirectional spatial smoothing covariance matrix is full rank.
Therefore, by following the standard MUSIC algorithm, the
angles of the multi-path channel components can be estimated.

Next we construct the receive beam beamforming matrix, which consists of the beam beamforming vectors that match the $L$ multipaths, respectively, as
\begin{equation}
		\small
\setlength\abovedisplayskip{1pt}
\setlength\belowdisplayskip{1pt}
  \mathbf{W}={{\left[ \frac{\mathbf{a}\left( {{\hat{\theta} }_{1}} \right)}{\left\| \mathbf{a}\left( {{\hat{\theta} }_{1}} \right) \right\|},\cdots ,\frac{\mathbf{a}\left( {{\hat{\theta} }_{L}} \right)}{\left\| \mathbf{a}\left( {{\hat{\theta} }_{L}} \right) \right\|} \right]}^{\mathrm{H}}} =\frac{1}{\sqrt{M}}{{\mathbf{A}}^{\mathrm{H}}\mathbf{\left( \hat{\Theta}  \right)}}\in {{\mathbb{C}}^{L\times M}}.
\end{equation}
Then the resulting signal at the BS can be written as
\begin{equation}
	\small
\begin{split}
\setlength\abovedisplayskip{1pt}
\setlength\belowdisplayskip{1pt}
   {{\mathbf{Y}}_{t,\mathrm{MP}}}&=\sqrt{{{P}_{t}}}\mathbf{Wh}{{\bm{\varphi }}^{\mathrm{H}}}+\mathbf{W}{{\mathbf{Z}}_{t}} \\
 & =\frac{\sqrt{{{P}_{t}}}}{\sqrt{M}}{{\mathbf{A}}^{\mathrm{H}}}\big( {\mathbf{\hat{\Theta }}} \big)\mathbf{A}\left( \mathbf{\Theta } \right)\bm{\alpha }{{\bm{\varphi }}^{\mathrm{H}}}+\frac{1}{\sqrt{M}}{{\mathbf{A}}^{\mathrm{H}}}\big( {\mathbf{\hat{\Theta }}} \big){{\mathbf{Z}}_{t}}.
\end{split}
\end{equation}
Similar to the analysis in Section \ref{los}, ${{\mathbf{Y}}_{t,\mathrm{MP}}}$  can be projected to the known pilot sequence $\bm{\varphi}$, which gives
\begin{equation}
\begin{split}
\setlength\abovedisplayskip{1pt}
\setlength\belowdisplayskip{1pt}
  {{\mathbf{y}}_{t,\mathrm{MP}}}& =\frac{1}{\rho }{{\mathbf{Y}}_{t,\mathrm{MP}}}\bm{\varphi }\\
  & =\frac{\sqrt{{{P}_{t}}}}{\sqrt{M}}{{\mathbf{A}}^{\mathrm{H}}}\big( {\mathbf{\hat{\Theta }}} \big)\mathbf{A}\left( \mathbf{\Theta } \right)\bm{\alpha }+\frac{1}{\sqrt{M{{\rho }^{2}}}}{{\mathbf{A}}^{\mathrm{H}}}\big( {\mathbf{\hat{\Theta }}} \big){{\mathbf{z}}_{t}'},
\end{split}
\end{equation}
where ${\mathbf{z}}_{t}'$ is i.i.d. CSCG noise with power ${\rho{\sigma }^{2}}$.
The LS estimation of $\bm{\alpha}$ can be written as
\begin{equation}
	\small
\begin{split}
\setlength\abovedisplayskip{1pt}
\setlength\belowdisplayskip{1pt}
\bm{\hat{\alpha }}& = \frac{{\sqrt M }}{{\sqrt {{P_t}} }}{({{\bf{A}}^{\rm{H}}}({\bf{\hat \Theta }}){\bf{A}}({\bf{\Theta }}))^{ - 1}}{{\bf{y}}_{t,{\rm{mp}}}}\\
& \mathop\approx \limits^{(b)} \frac{{\sqrt M }}{{\sqrt {{P_t}} }}{({{\bf{A}}^{\rm{H}}}({\bf{\hat \Theta }}){\bf{A}}({\bf{\hat \Theta }}))^{ - 1}}{{\bf{y}}_{t,{\rm{mp}}}}
,\label{bmalpha}
\end{split}
\end{equation}
where $(b)$ is based on ${\mathbf{\Theta }}\approx{\mathbf{\hat{\Theta }}}$.
Then the estimation of $\mathbf{h}_{\mathrm{lp}}$ can be formulated as ${{\mathbf{\hat{h}}}_{\mathrm{lp}}}=\mathbf{A}\left( {\mathbf{\hat{\Theta }}} \right)\bm{\hat{\alpha }}$, and the estimation error of ${{\mathbf{{h}}}_{\mathrm{lp}}}$ can be written as
\begin{equation}
		\small
\begin{split}
\setlength\abovedisplayskip{1pt}
\setlength\belowdisplayskip{1pt}
  & {{{\mathbf{\tilde{h}}}}_{\mathrm{lp}}}={{{\mathbf{\hat{h}}}}_{\mathrm{lp}}}-\mathbf{h}_{\mathrm{lp}}=\mathbf{A}\big( {\mathbf{\hat{\Theta }}} \big)\bm{\hat{\alpha} }-\mathbf{A}\left( \mathbf{\Theta } \right)\bm{\alpha }  \\
 & =\frac{\sqrt{M}}{\sqrt{{{P}_{t}}}}\mathbf{A}\big( {\mathbf{\hat{\Theta }}} \big){{\left( {{\mathbf{A}}^{\mathrm{H}}}\big( {\mathbf{\hat{\Theta }}} \big)\mathbf{A}\big( {\mathbf{\hat{\Theta }}} \big) \right)}^{-1}}{{\mathbf{y}}_{t,\mathrm{MP}}}-\mathbf{A}\big( \mathbf{\Theta } \big)\bm{\alpha }.\label{tildehulpmulti}
\end{split}
\end{equation}
When the angle is perfectly estimated, \eqref{tildehulpmulti} can be written as
\begin{equation}
		\small
{{{\mathbf{\tilde{h}}}_{\mathrm{lp}}}=\frac{1}{\sqrt{{{P}_{t}}{\rho}^{2}}}\mathbf{A}\big( {\mathbf{\hat{\Theta }}} \big){{\Big( {{\mathbf{A}}^{\mathrm{H}}}\big( {\mathbf{\hat{\Theta }}} \big)\mathbf{A}\big( {\mathbf{\hat{\Theta }}} \big) \Big)}^{-1}}{{\mathbf{A}}^{\mathrm{H}}}\big( {\mathbf{\hat{\Theta }}} \big){{\mathbf{z}}_{t}'}}.
\end{equation}
Then the MMSE of ${\mathbf{h}}$ can be written as
\begin{equation}
 {{{e}_{\mathrm{lp}}}=\mathbb{E}\big[ {{\big\| {{{\mathbf{\tilde{h}}}}_{\mathrm{lp}}} \big\|}^{2}} \big]=\frac{L{{\sigma }^{2}}}{{{P}_{t}\rho}}}.
\end{equation}
The proofs are omitted for brevity.
Compared with $e_\mathrm{pc}$ in \eqref{RMSEe1}, it can be shown that when $L<M$, the MMSE of the little pilot channel estimation scheme is smaller than that using pilot-consuming channel estimation method.

In order to further detect the communication symbol, the
receive beamforming matrix ${{\mathbf{v}}_{\mathrm{lp}}}=\frac{{{{\mathbf{\hat{h}}}}_{\mathrm{lp}}}}{\left\| {{{\mathbf{\hat{h}}}}_{\mathrm{lp}}} \right\|}$ is used, and the resulting
signal is written as
\begin{equation}
{{y_{d,\mathrm{lp}}(n)}=\mathbf{v}_{\mathrm{lp}}^{\mathrm{H}}{{\mathbf{y}}_{d}(n)}
=\sqrt{{{P}_{d}}}\mathbf{v}_{\mathrm{lp}}^{\mathrm{H}}\mathbf{h}{s}\left( n \right)+{z}_{d,\mathrm{lp}}\left( n \right)},
\end{equation}
where ${z}_{d,\mathrm{lp}}\left( n \right)=\mathbf{v}_{\mathrm{lp}}^{\mathrm{H}}{{\mathbf{z}}_{d}\left( n \right)}$.

\emph{Theorem 3:}
The expected SNR of the signal in \eqref{afterbeamangleL1} can be written as
\begin{equation}
  {{\gamma }_{\mathrm{lp}}}\!=\!\mathbb{E}\bigg[\! \frac{{{P}_{d}}{{\big| \mathbf{v}_{\mathrm{lp}}^{\mathrm{H}}\mathbf{h} \big|}^{2}}}{{{\sigma }^{2}}} \!\bigg]\!=\!\frac{{{P}_{d}}}{{{\sigma }^{2}}}{{\mathbf{h}}^{\mathrm{H}}}\mathbb{E}\left[\! {{\mathbf{v}}_{\mathrm{lp}}}\mathbf{v}_{\mathrm{lp}}^{\mathrm{H}} \!\right]\mathbf{h}.\label{gammaulpmulti}
\end{equation}
The upper bound of the expected SNR can be written as
\begin{equation}
  {{{\gamma }_{\mathrm{lp}}} \le {{\gamma }_{\mathrm{lp,upper}}}=\frac{{{P}_{d}}}{{{\sigma }^{2}}}{{\left\| \mathbf{h} \right\|}^{2}}}.\label{SNRmultipathUP}
\end{equation}
The proofs are omitted for brevity.

\section{Simulation Results}
In this section, simulation results are provided to compare
the performance of the conventional channel estimation method and the proposed ISSAC-based channel estimation method.
The number of all path components is set as $L=3$.
The path gains $\alpha_{l}$ are generated as following:
$\alpha_{l}\sim\mathcal{C}\mathcal{N}\left( 0, 1\right)$.
Define ${{\bar{P}}_{t}}=\frac{{{P}_{t}}}{{{\sigma }^{2}}}$ and
${{\bar{P}}_{d}}=\frac{{{P}_{d}}}{{{\sigma }^{2}}}$ as the transmit signal-to-noise ratio
(SNR) by the UE in pilot and data transmission phases, respectively.
Unless otherwise stated, we set $\rho=3, {\bar{P}}_{t}={\bar{P}}_{d}=-10$dB, and the number of BS antennas is $M=32$.
To achieve the angle information, we use MUSIC algorithm for high resolution.

\begin{figure}[htbp]
  \centering
    \begin{subfigure}{0.35\textwidth}
      \centering
      \includegraphics[width=1\linewidth]{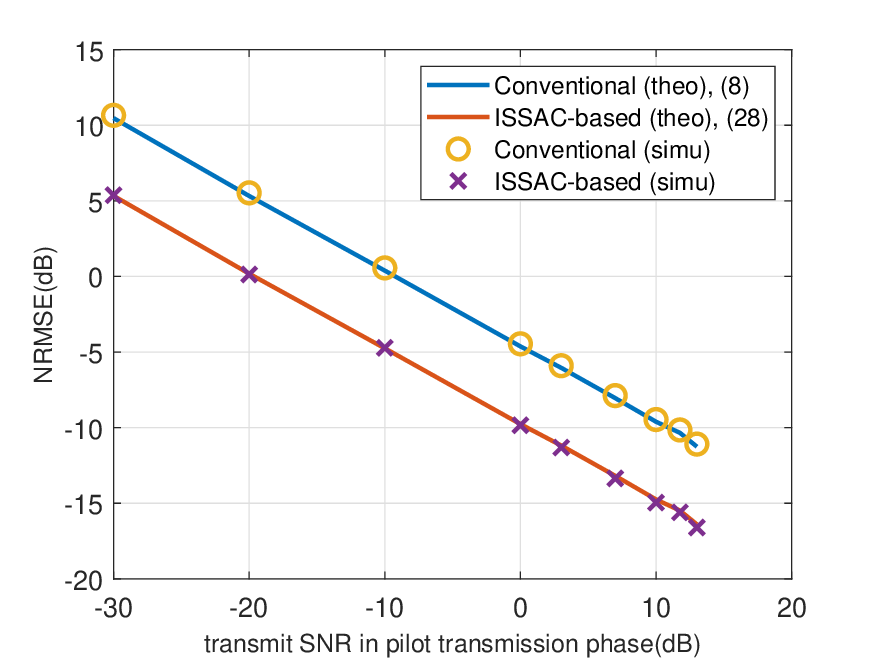}
        \caption{NRMSE versus transmit SNR in pilot transmission phase ${\bar{P}}_{t}$.}
        \label{NRMSEvsPt}
    \end{subfigure}   
     \begin{subfigure}{0.35\textwidth}
      \centering
      \includegraphics[width=1\linewidth]{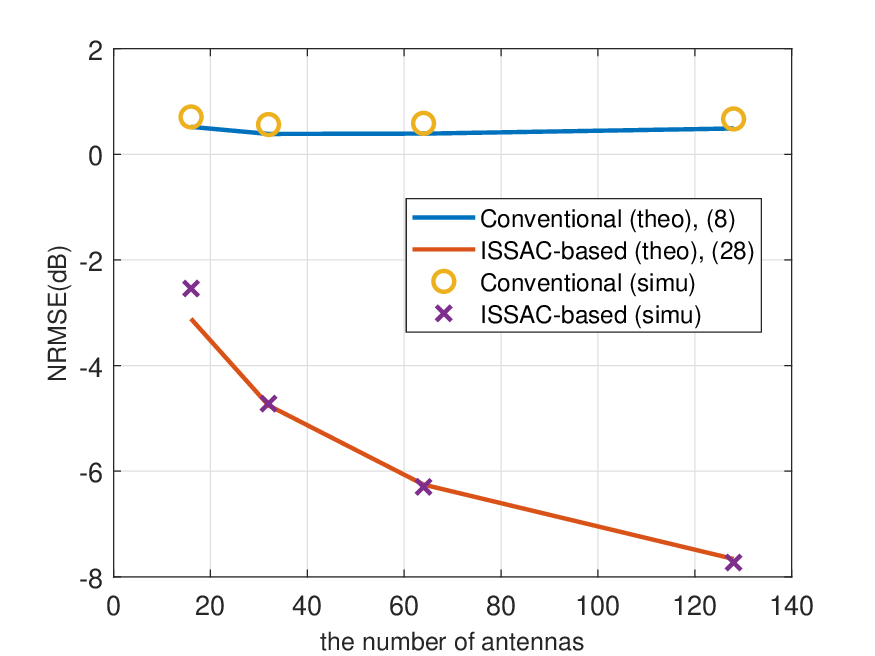}
        \caption{NRMSE versus the number of antennas $M$.}
        \label{NRMSEvsM}
    \end{subfigure}
\caption{NRMSE comparison of two methods.}\label{fig:totalNRMSE}
 \vspace{-0.1cm}
\end{figure}

Fig. \ref{NRMSEvsPt} and Fig. \ref{NRMSEvsM} plot the channel error $e_\mathrm{cp}$ and $e_\mathrm{lp,MP}$ and their theoretical value versus ${\bar{P}}_{t}$ and the number of antennas, respectively.
It can be observed that the theoretical values of both methods perfectly match with the actual values.
Specifically, it is shown that the channel error is independent of the number of antennas for the conventional pilot-consuming channel estimation method,
while for the other method, it decreases as the increase of the number of antennas.
Besides, the normalized root mean squared error (NRMSE) of the channel estimation method based on ISSAC
is much smaller than the conventional channel estimation method, which is consistent with our theoretical analysis.

\begin{figure}[!t]
 \setlength{\belowcaptionskip}{-0.2cm}
  \centering
  \centerline{\includegraphics[width=2.6in,height=1.8in]{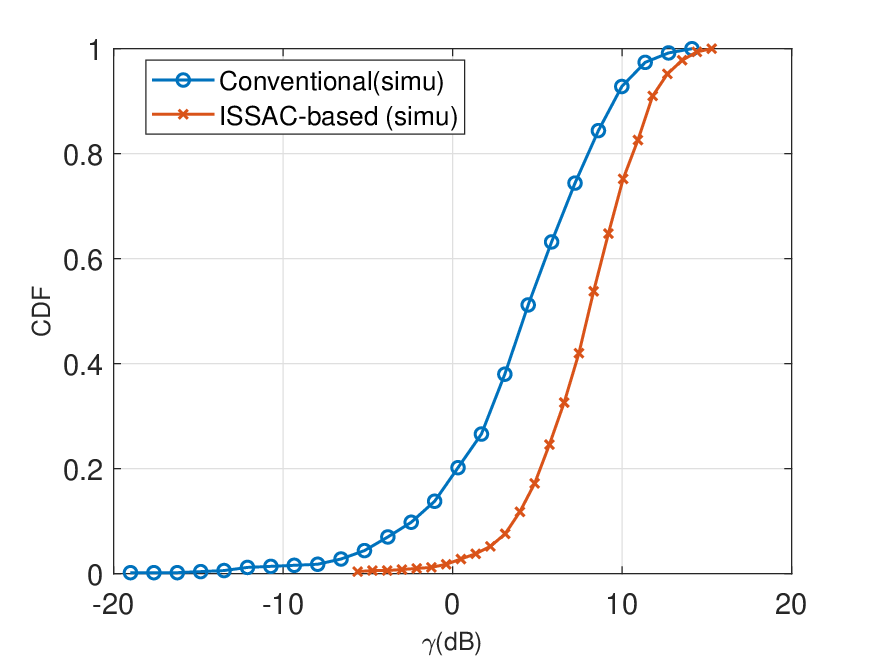}}
  \caption{CDF of receive SNR.}
  \label{CDF}
   \vspace{-0.3cm}
  \end{figure}

Fig. \ref{CDF} plots the cumulative distribution function (CDF) of $\gamma_\mathrm{cp}$ and $\gamma_\mathrm{lp}$ when decoding data signals in \eqref{gamma1initial} and \eqref{gammaulpmulti}.
It can be observed that $\gamma_\mathrm{lp}$ is larger than $\gamma_\mathrm{cp}$ with the consideration of 90-percentile SNR,
which indicates that the SNR performance of data transmission using channel estimation based on ISSAC is much better than that of the conventional channel estimation method.

\begin{figure}[htbp]
  \centering
    \begin{subfigure}{0.35\textwidth}
      \centering
      \includegraphics[width=1\linewidth]{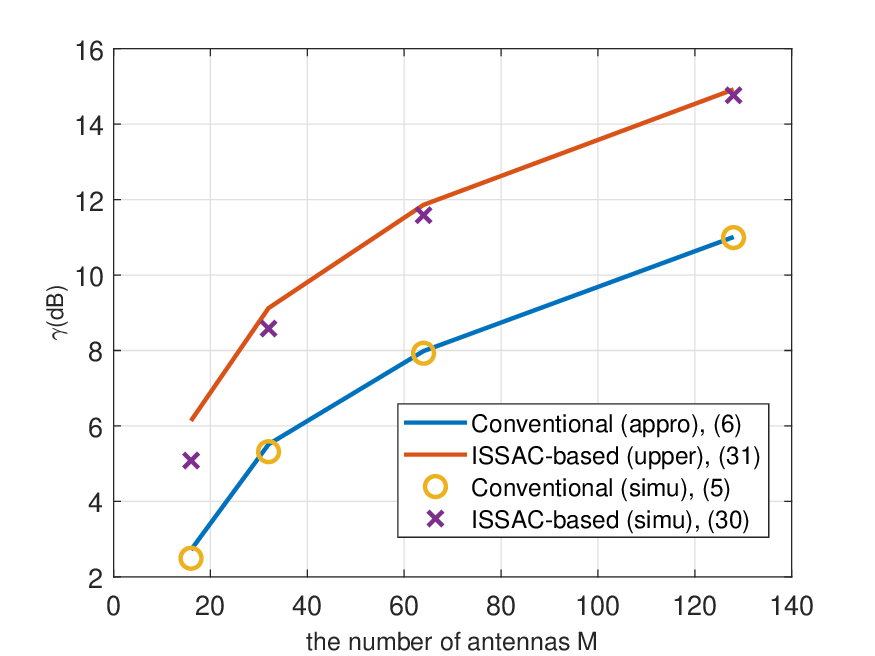}
        \caption{Receive SNR versus the number of antennas $M$ in the phase of data transmission.}
        \label{gammavsM}
    \end{subfigure}   
     \begin{subfigure}{0.35\textwidth}
      \centering
      \includegraphics[width=1\linewidth]{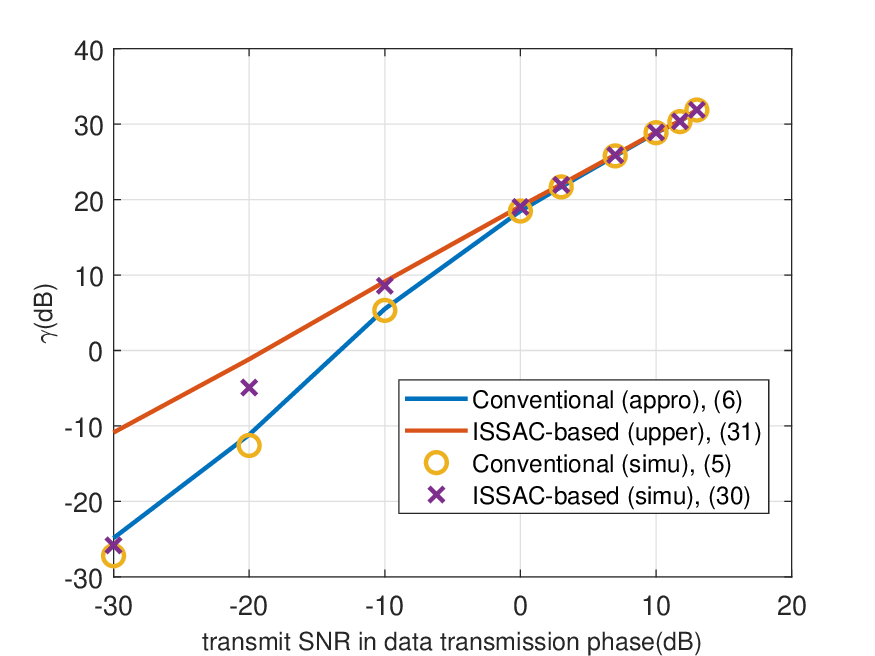}
        \caption{Receive SNR versus transmit SNR ${\bar{P}}_{d}$ in the phase of data transmission.}
        \label{gammavsPd}
    \end{subfigure}
\caption{Receive SNR comparison of two methods.}\label{fig:totalSNR}
 \vspace{-0.1cm}
\end{figure}

Fig. \ref{gammavsM} and Fig. \ref{gammavsPd} plot the receive SNR when decoding data signals using two methods in \eqref{gamma1initial} and \eqref{gammaulpmulti} and their approximation value and upper value in \eqref{gamma1app} and \eqref{SNRmultipathUP} respectively versus the $M$ and ${\bar{P}}_{d}$, respectively.
It can be observed from Fig. \ref{gammavsM} that the approximation value and upper value in \eqref{gamma1app} and \eqref{SNRmultipathUP} are very close to the exact values in \eqref{gamma1initial} and \eqref{gammaulpmulti}.
It is observed in Fig. \ref{gammavsPd} that within a certain range, e.g., roughly -30dB to 5dB, the receive SNR for the method we proposed is much higher than that of the conventional channel estimation method,
and this gap is widest at around -20dB.
This indicates that the little pilot scheme we proposed can provide more accurate channel estimation under poor SNR.

\section{Conclusion}
In this paper, we propose an ISSAC-based channel estimation method, which requires little or even no pilot,
yet still achieves accurate CSI estimation.
For the proposed method, we first estimate the angles of the multi-path channel components in a pilot-free manner during the phase of data transmission.
Then, by using very little pilots, the multi-path channel coefficients are further estimated with the beamforming gain.
The basic LoS channel is first considered
to analyze the MMSE of channel estimation and the resulting beamforming gains,
which indicates that our proposed method significantly outperforms the conventional methods purely based on channel training and obtain higher SNR without pilots.
Then the study is extended to the more general multipath channels.
Simulation results are provided to demonstrate our theoretical results.
\section{Acknowledgment}
This work was supported by the National Natural Science Foundation of China with grant number 62071114.
\begin{appendices}
\section{Proof of Theorem 1}
The derivation of the closed-form expression of \eqref{gamma1initial} is non-trivial.
To gain more insights, we approximate $\mathbb{E}\left[ {{\mathbf{v}}_{\mathrm{pc}}}\mathbf{v}_{\mathrm{pc}}^{\mathrm{H}} \right]$ in \eqref{gamma1initial} as
\begin{small}
\begin{equation}
\begin{split}
\setlength\abovedisplayskip{2pt}
\setlength\belowdisplayskip{2pt}
  & \mathbb{E}\left[ {{\mathbf{v}}_{\mathrm{pc}}}\mathbf{v}_{\mathrm{pc}}^{\mathrm{H}} \right] =\mathbb{E}\bigg[ \frac{{{{\mathbf{\hat{h}}}}_{\mathrm{pc}}}\mathbf{\hat{h}}_{\mathrm{pc}}^{\mathrm{H}}}{{{\big\| {{{\mathbf{\hat{h}}}}_{\mathrm{pc}}} \big\|}^{2}}} \bigg]\approx \frac{\mathbb{E}\left[ {{{\mathbf{\hat{h}}}}_{\mathrm{pc}}}\mathbf{\hat{h}}_{\mathrm{pc}}^{\mathrm{H}} \right]}{\mathbb{E}\big[ {{\big\| {{{\mathbf{\hat{h}}}}_{\mathrm{pc}}} \big\|}^{2}} \big]} \\
 & =\frac{\mathbb{E}\left[ \left( \mathbf{h}+{{{\mathbf{\tilde{h}}}}_{\mathrm{pc}}} \right){{\left( \mathbf{h}+{{{\mathbf{\tilde{h}}}}_{\mathrm{pc}}} \right)}^{\mathrm{H}}} \right]}{\mathbb{E}\left[ {{\left\| \mathbf{h}+{{{\mathbf{\tilde{h}}}}_{\mathrm{pc}}} \right\|}^{2}} \right]} =\frac{\mathbf{h}{{\mathbf{h}}^{\mathrm{H}}}+\frac{{{\sigma }^{2}}}{{{P}_{t}}\rho }{{\mathbf{I}}_{M}}}{{{\left\| \mathbf{h} \right\|}^{2}}+\frac{M{{\sigma }^{2}}}{{{P}_{t}}\rho }},\label{EvvH1}
\end{split}
\end{equation}
\end{small}where the approximation holds for $M\gg 1$.
According to \eqref{EvvH1}, the expected SNR of the signal in \eqref{gamma1initial} can be further approximated as
\begin{small}
\begin{equation}
\begin{split}
\setlength\abovedisplayskip{2pt}
\setlength\belowdisplayskip{2pt}
  {{\gamma }_{\mathrm{pc}}}& \approx\frac{{{P}_{d}}{{\left\| \mathbf{h} \right\|}^{2}}\left( \frac{{{P}_{t}}\rho }{M{{\sigma }^{2}}}{{\left\| \mathbf{h} \right\|}^{2}}+\frac{1}{M} \right)}{{{\sigma }^{2}}\left( \frac{{{P}_{t}}\rho }{M{{\sigma }^{2}}}{{\left\| \mathbf{h} \right\|}^{2}}+1 \right)} \\
 & =\frac{{{P}_{d}}{{\left\| \mathbf{h} \right\|}^{2}}\left( \rho \mathrm{SNR}_{t}+\frac{1}{M} \right)}{{{\sigma }^{2}}\left( \rho \mathrm{SNR}_{t}+1 \right)}=\frac{{{P}_{d}}{{\left\| \mathbf{h} \right\|}^{2}}}{{{\sigma }^{2}}}\left( 1-\frac{1-\frac{1}{M}}{\rho \mathrm{SNR}_{t}+1} \right).
\end{split}
\end{equation}
\end{small}
\end{appendices}

\end{document}